\documentstyle[12pt]{article}
\begin{document}

\begin{center}
{\bf
ANALYTIC FORM FACTORS OF HYDROGENLIKE ATOMS FOR DISCRETE-CONTINUUM TRANSITIONS.
II. GENERAL CASE\\}
\vspace*{1cm}
O.Voskresenskaya$^{1,2,3}$\footnote{E-mail:
Olga.Voskresenskaja@mpi-hd.mpg.de; voskr@cv.jinr.dubna.su}\\
\vspace*{.25cm}
{\it$^1$ Institut f\"ur Theoretische Physik der Universit\"at,
69120 Heidelberg, Germany\\
\it$^2$ Max-Planck-Institut f\"ur Kernphysik, Postfach 103980,
69029 Heidelberg, Germany\\
$^3$ Joint Institute for Nuclear Research,
141980 Dubna, Moscow Region, Russia\\}
\end{center}

\vspace*{.25cm}
\begin{abstract}
{\small A general formula for bound-continuous transition form factors
is derived. It is shown that these form factors can be represented
in the form of finite sum of terms with simple analytical structure.}

\end{abstract}

\vspace*{.5cm}
In the previous paper \cite{voskr1} it have been shown that the form factors
of transitions from $nS\; (n00)$ - states of hydrogenlike atoms \cite{faust}
to the state of continuous spectra with definite value of relative momenta
$\vec p$ may be expressed in the terms of the classical polynomials
in a rather simple way.
Below this result is generalized for the case of  transition
from arbitrary initial bound states.

The transition form factors are defined as follows:
\begin{equation}
S_{fi}(\vec q)=\int \psi_f^{\ast}(\vec r)e^{i\vec q\vec r}
\psi_i(\vec r)d^3r
\label{e1}\, ,
\end{equation}
Here, $\psi_{i(f)}$ are the wave functions of initial (final) states.

According to \cite{sommer} (see also \cite{land}), the final state wave
function must be choose in the form
\begin{equation}
\psi_i(\vec r)\equiv\psi_{n00}(\vec r)
\label{e2}\, .
\end{equation}
For arbitrary initial bound state
\begin{equation}
\psi_i(\vec r)=\psi_{nlm}(\vec r)=
Y_{lm}\left(\frac{\vec r}{r}\right)R_{nl}(r)\,,
\label{351}
\end{equation}
\begin{eqnarray}
R_{nl}(r)&=&\frac{2\omega^{\frac{3}{2}}}{\Gamma(2l+2)}
\left[\frac{\Gamma(n+l+1)}{n\Gamma(n-l)}\right]^{\frac{1}{2}}
\cdot(2\omega r)^l
\cdot\Phi(-n+l+1,2l+2;2\omega r)\cdot \exp(-\omega r)\nonumber
\end{eqnarray}
$$~~~~=2\omega^{\frac{3}{2}}
\left[\frac{\Gamma(n-l)}{n\Gamma(n+l+1)}\right]^{\frac{1}{2}}
\cdot(2\omega r)^l
\cdot L^{2l+1}_{n-l-1}(2\omega r)\cdot \exp(-\omega r)\,,$$
$$
\omega=\mu\alpha/n\,,
$$
where $L^{2l+1}_{n-l-1}$ are the associated Laguerre polynomials.

\newpage
Making use of the recurrence relations \cite{ryzh}
\begin{equation}
L_k^{\lambda+1}(x)=\frac{1}{x}\left[(k+\lambda+1)L_{k-1}^{\lambda}(x)
-(k+1)L_{k}^{\lambda}(x)\right]
\label{e4}
\end{equation}
and the representation of the Laguerre polynomials
in terms of the generating function
\begin{equation}
L_k^{\lambda}(x)=\Delta_z^{(k)}\left[(1-z)^{-(\lambda+1)}\exp
\left(\frac{xz}{z-1}\right)\right]
\label{e5}\, ,
\end{equation}
where operator $\Delta_z^{(k)}$ is defined as follows
\begin{equation}
\Delta_z^{(k)}\left[f(z)\right]=\frac{1}{k!}
\left.\left(\frac{d^k}{dz^k}f(z)\right)\right\vert_{z=0}
\label{e6}\, ,
\end{equation}
let us rewrite the radial part of initial state wave
function in the form
\begin{equation}
R_{nl}=\frac{\omega^{\frac{1}{2}}}{r}
\left[\frac{\Gamma(n-l)}{n\Gamma(n+l+1)}\right]^{\frac{1}{2}}
\cdot(2\omega r)^l
\cdot\left[(n+l)\Delta_z^{(n-l-2)}-(n-l)\Delta_z^{(n-l-1)}\right]
\label{e7}
\end{equation}
$$\times\left[(1-z)^{-(l+1)}\exp\left(-\omega(z)r\right)\right]\,,$$
\begin{equation}
\omega(z)=\omega\cdot(1+z)(1-z)^{-1}
\label{e11}
\end{equation}
more convenient for the further calculations.

Then it is not difficult to see that transition form factors
(\ref{e1}) may be represent as a linear combination of the quantities
\begin{equation}
I_{lm}^{j}=\Delta_z^{(j)}\left[(1-z)^{-(2l+1)}
J_{lm}(\vec q,\vec p,z)\right]
\label{eq:352}\, ,
\end{equation}
\begin{eqnarray}
J_{lm}(\vec q,\vec p,z)&=&\int \frac{d^3r}{r}Y_{lm}
\left(\frac{\vec r}{r}\right)
\Phi\left[i\xi,1;i(pr+\vec p\vec r)\right]
\label{eq:353}
\end{eqnarray}
$$\times\exp[i(\vec q-\vec p)\vec r-\omega(z)\cdot r]\cdot(2\omega r)^l\cdot
\exp\left[-\omega(z)r\right]\,,$$
$$j=n-l-2,\;n-l-1\,.$$

In order to calculate (\ref{eq:353}), it is useful to represent the
hypergeometrical function in (\ref{351}) in the form
\begin{eqnarray}
\Phi\left[i\xi,1;i(pr+\vec p\vec r)\right]&=&-\frac{1}{2\pi i}
\oint\limits_{C}^{} (-t)^{i\xi-1}(1-t)^{-i\xi}
\cdot\exp[i\cdot t(pr+\vec p\vec r)]dt\,.
\label{eq:354}
\end{eqnarray}

Using the following relations
\begin{equation}
\label{eq:355}
\exp(i\vec\tau\vec r)=4\pi\sum_{l=0}^{\infty}\sum_{m=-l}^{l}
i^{l}Y_{lm}\left(\frac{\vec \tau}{\tau}\right)
Y^{\ast}_{lm}\left(\frac{\vec r}{r}\right)j_{l}(\tau r)\,,
\end{equation}
\begin{equation}
\label{eq:356}
j_{l}(x)=\sqrt{\frac{\pi}{2x}}J_{l+\frac{1}{2}}(x)\,,
\end{equation}
\begin{equation}
\label{eq:357}
\int\limits_{0}^{\infty}r^{l+\frac{1}{2}}J_{l+\frac{1}{2}}(\tau r)
e^{-\bar\omega\cdot r}dr=
\frac{(2\tau)^{l+\frac{1}{2}}\Gamma(l+1)}
{\sqrt{\pi}(\tau^2+\bar \omega^2)^{l+1}}\,,
\end{equation}
where
\begin{equation}
\label{eq:358}
\vec\tau=\vec q-\vec p(1-t),\quad \bar\omega=\omega(z)-ip\cdot t\,,
\end{equation}
after simple calculations we find
\begin{eqnarray}
J_{lm}(\vec q,\vec p,z)&=&-\frac{\Gamma(l+1)}{2\pi i}
\oint\limits_{C}^{ }dt(-t)^{i\xi-1}(1-t)^{-i\xi}
\cdot\frac{4\pi(4i\omega)^lY_{lm}\left(\vec \tau/\tau\right)\tau^l}
{(\tau^2+\bar \omega^2)^{l+1}}\,.
\label{eq:359}
\end{eqnarray}

It is easy to check that
$$\tau^2+\bar \omega^2=a(1-t)+c\cdot t\,,$$
\begin{equation}
\label{eq:362}
a=\omega^2(z)+\vec\Delta^2\,,\quad c=\left[\omega(z)-ip\right]^2+q^2\,.
\end{equation}

Further, according to \cite{war}, we get
\begin{eqnarray}
\label{eq:360}
Y_{lm}\left(\frac{\vec \tau}{\tau}\right)\tau^l
=\sum_{l_1=0}^{l}q^{l_1}(-p)^{l-l_1}
\end{eqnarray}
$$\times\left[\frac{4\pi\Gamma(2l+2)}
{\Gamma(2l_1+2)\Gamma(2l-2l_1+2)}\right]^{\frac{1}{2}}
\cdot (1-t)^{l-l_1}\left[Y_{l_1}\left(\frac{\vec q}{q}\right)\otimes
Y_{l-l_1}\left(\frac{\vec p}{p}\right)\right]_{lm}\,,$$
\begin{eqnarray}
\label{eq:361}
\left[Y_{l_1}\left(\frac{\vec q}{q}\right)\otimes
Y_{l-l_1}\left(\frac{\vec p}{p}\right)\right]_{lm}
=\sum_{m_1+m_2=m}C^{lm}_{l_1m_1(l-l_1)m_2}
\cdot Y_{l_1m_1}\left(\frac{\vec q}{q}\right)\cdot
Y_{(l-l_1)m_2}\left(\frac{\vec p}{p}\right)\,.
\end{eqnarray}
Taking into account (\ref{eq:362}) and (\ref{eq:360}), it is easy
to see that (\ref{eq:359}) is the superposition of the quantities
\begin{equation}
-\frac{1}{2\pi i}\oint\limits_{C}^{} \frac{t^{i\xi-1}(1-t)^{-i\xi+l-l_1}}
{[a(1-t)+ct]^{l+1}}
\end{equation}
\begin{eqnarray}
&&=a^{-(l+1)}\frac{\Gamma(1-i\xi+l-l_1)}{\Gamma(1-i\xi)}
F\left(i\xi,l+1;l-l_1+1;1-c/a\right)\nonumber\\
&&=a^{i\xi-l-1}c^{-i\xi}\frac{\Gamma(1-i\xi+l-l_1)}
{\Gamma(1-i\xi)}
F\left(i\xi,-l_1;l-l_1+1;1-a/c\right)\nonumber\\
&&=a^{i\xi-l-1}c^{-i\xi}\frac{\Gamma(l-l_1+1)\Gamma(l+1-i\xi)}
{\Gamma(l+1)\Gamma(1-i\xi)}F(i\xi,-l_1;i\xi-l;a/c)\nonumber\\
&&=\sum_{s=0}^{l_1}(-1)^{l-s}
\frac{\Gamma(i\xi+s)\Gamma(l_1+1)}
{\Gamma(l_1-s+1)\Gamma(i\xi-l+s)\Gamma(s+1)\Gamma(l+1)}
a^{i\xi+s-l-1}c^{-s-i\xi}\nonumber\\
&&=(1-z)^{2l+2}\sum_{s=0}^{l_1}(-1)^{l-s}
\frac{\Gamma(i\xi+s)\Gamma(l_1+1)}
{\Gamma(l_1-s+1)\Gamma(i\xi-l+s)\Gamma(s+1)\Gamma(l+1)}\nonumber
\label{eq:363}
\end{eqnarray}
$$\times D_1^{i\xi+s-l-1}D_2^{-s-i\xi}\,,$$
where $D_{1,2}$ are defined as follows:
\begin{equation}
D_1=(1+z^2)(\omega^2+\vec\Delta^2)-2z(\vec\Delta^2-\omega^2)\,,
\label{e3}
\end{equation}
$$D_2=(\omega-ip)^2+q^2-2z(q^2-p^2-\omega^2)+z^2[(\omega+ip)^2+q^2]\,.$$

The further calculations are the same as in \cite{voskr1}.

Omitting the simple but cumbersome algebra, let us present the final
expression for transition form factors:
\begin{eqnarray}
S_{\vec p,nlm}(\vec q)&=&4\pi\cdot 2^{2l}i^l\omega^{l+\frac{1}{2}}
\left[\frac{\Gamma(n-l)}{n\Gamma(n+l+1)}\right]^{\frac{1}{2}}\nonumber\\
\label{eq:364}
\end{eqnarray}
$$
\times\sum_{s=0}^{l}G_{lms}(\vec p,\vec q)H_{nls}(\vec p,\vec q)
\cdot(\omega^2+\Delta ^2)^{i\xi+s-l-1}[(\omega-ip)^2+q^2]^{-s-i\xi}\,;
$$
\begin{eqnarray}
\label{eq:365}
G_{lms}(\vec p,\vec q)&=&(-1)^{l-s}
\frac{\Gamma(i\xi+s)}{\Gamma(i\xi-l+s)\Gamma(s+1)}
\end{eqnarray}
$$\times\sum_{l_1=s}^{l}
\left[\frac{4\pi\Gamma(2l+2)}{\Gamma(2l_1+2)\Gamma(2l-2l_1+2)}\right]^
{\frac{1}{2}}
\cdot\frac{\Gamma(l_1+1)}{\Gamma(l_1-s+1)}q^{l_1}(-p)^{l-l_1}$$
$$\times\left[Y_{l_1}\left(\frac{\vec q}{q}\right)\otimes
Y_{l-l_1}\left(\frac{\vec p}{p}\right)\right]_{lm}\,;$$
\begin{eqnarray}
\label{eq:366}
H_{nls}(\vec p,\vec q)&=&(n+l)F_{n_1ls}(\vec p,\vec q)-
(n-l)F_{n_2ls}(\vec p,\vec q)\,;
\end{eqnarray}
$$n_1=n-l-1\,,\quad n_2=n-l-2\,;$$
\begin{eqnarray}
\label{eq:367}
F_{n_{1(2)}ls}(\vec p,\vec q)&=&
\frac{\Gamma(l-s+\frac{1}{2}-i\xi)}{\Gamma(2l-2s+1-2i\xi)}
\sum_{k=0}^{n_{1(2)}}w^kC_k^{(i\xi+s)}(v)
\end{eqnarray}

$$\times\frac{\Gamma(n_{1(2)}-k+2l-2s+1-2i\xi)}
{\Gamma(n_{1(2)}-k+l-s+\frac{1}{2}-i\xi)}
\cdot P_{n_{1(2)}-k}^{(l-s-\frac{1}{2}-i\xi,l-s+\frac{1}{2}-i\xi)}(u)\,.$$\\

Thus, the form factors for transition from arbitrary bound states of
hydrogenlike atoms to the ``$\vec p-state$''
of continuous spectra are represented as the superposition of finite
number of terms with simple analytical structure and can be evaluated
numerically with arbitrary degree of accuracy.

Eqs. (\ref{eq:364})-(\ref{eq:367}) are the generalization of the
results of \cite{MASS33,voskr1}.
\vspace*{.5cm}
\begin{center}

{\large \bf Acknowledgments}

\end{center}

\vspace*{.5cm}

I would like to thank the Institute for Theoretical Physics
at Heidelberg University and the Max-Planck-Institut f\"ur Kernphysik,
where this work was carried out, for the hospitality.
I also thank Alexander Tarasov for stimulating discussions.

\end{document}